# New 3D multipass amplifier based on Nd:YAG or Nd:YVO$_4$ crystals

Sebastien Forget[1](✉), François Balembois[1], Patrick Georges[1], Pierre-Jean Devilder[2]

[1] *Laboratoire Charles Fabry de l'Institut d'Optique, UMR 8831 du CNRS, Université Paris-Sud, 91403 Orsay, France*

[2] *JDS Uniphase, 31 chemin du vieux chêne, 38941 Meylan, France*

ABSTRACT :

We report on new simple and efficient multipass amplifiers using prisms or corner cubes to perform several passes in different planes of incidence. This scheme provides an optimized overlap between the signal passes and the pumped volume. We investigated our amplification geometry with Nd:YAG and Nd: YVO$_4$ crystals : the use of a low doped (0.3%) Nd:YVO$_4$ crystal allowed better thermal behaviour and higher performances. We amplified a pulsed microlaser (110 mW of average power at 1064 nm) and obtained a diffraction-limited output beam with an average power of 5.7 W for 15 W of pump power and a small-signal gain of 56 dB in a 6-pass configuration.

## 1 Introduction

Passively Q-switched microchip lasers are very simple, compact and reliable sources that can provide single mode, high repetition rate, near infrared and sub-nanosecond pulses with diffraction limited output [1-3]. They are consequently of strong interest for many applications : thanks to high peak power pulses, efficient conversion from 1064 nm to UV (355 nm) can be obtained in non linear crystals. However the available average power in the UV is limited to a few tens of milliwatts, which is not enough for applications in the material processing domain such as integrated circuit repair or rapid prototyping: in those case several watts of near infrared radiation are needed to reach several hundreds of milliwatts in the UV. A very fruitful approach to increase the energy per pulse is the use of a MOPA (Master Oscillator Power Amplifier) design with a microlaser oscillator and an amplifier [4]. In order to raise the microlaser output up to multiwatt level, the design of the amplifier is then critical : it has to provide both high gain and efficient energy extraction in the same time. A solution widely used is to make the laser beam pass several times into the amplifier[5-10] : the first passes allow strong amplification whereas the last passes extract the energy stored in the crystal. Various methods have been described in the literature: a first classical arrangement is to form an optical resonator in which the number of passes could be large (on the order of 100). Such regenerative amplifiers [5] produce large total gain but are relatively complex and expensive since active elements have to be used in the cavity to extract the signal. Another way to achieve multipass amplification is to make the signal beam pass several times through the crystal with geometrical path slightly different each time [6-10]. The number of pass is reduced (typically 10) and this configuration is much less complicated than the previous one. In the case of Nd-doped crystals, the geometrical multipass allows good extraction efficiency for relatively high input power. For example, Plaessman *et al* [6] obtained, with a 14-pass confocal planar geometry and a Nd:YAG crystal, an extracted power of 2.3 W and a gain of about 7 dB for 13W of pump power. However, the geometrical multipass amplifiers described in the literature are for the most part planar, resulting in a non-optimal overlap between the gain volume and the laser beams. Planar multipass geometry consequently reduces the available gain, leading to an increase of the number of passes necessary to reach optimal performances. Some tri-dimensionnal schemes, where the signal beams travel in different planes of incidence, were described [9-10] but were inserted in much more complex structures and lacked of either simplicity or compactness. In this paper we describe a novel simple and compact 3D architecture that allows optimal overlap between the different passes in the gain medium and the pump volume, leading to a very efficient extraction of the energy stored in the crystal. We experimented our multipass system with Nd:YAG Nd:YVO$_4$ crystals.

## 2 Experimental setup

The principle of our 3D scheme can be compared with the so-called thin disk structure [11, 12], but here the pump and signal are exchanged. The structure used is shown on figure 1 : the signal beam from a microlaser is focused in the gain medium through a 1064 nm AR coated lens L. We used a doublet to minimize spherical aberrations. A dichroïc plane mirror M coated on the rear face of the crystal reflects the beam back towards the doublet. At this time the beam has made two passes in the gain (in a plane marked by the path "1-2" on figure 1). An AR-coated right-angle prism then allows a lateral shift of the beam, that is, the laser beam now travels in another plane of incidence. The beam is then redirected toward the crystal through the same lens L (pass "3" on figure 1). After another reflexion on the mirror M, the signal beam has traveled 4 times in the crystal in two different planes, namely 1-2 and 3-4. This path is again repeated 2 times with two other prisms such leading to a total of eight passes in the gain medium in four different planes of incidence (see figure 1b).

The injected laser beam to be amplified was given by a microlaser from JDS Uniphase (High Power Series Nanolaser), providing 110 mW of average power at a wavelength of 1064 nm. The pulse duration was about 800 ps and the repetition rate was 28 kHz. The pumping geometry is described on figure 2 : the crystal was pumped by a fiber coupled laser diode providing up to 15 W CW at 808 nm. The core diameter of the fiber was 250 µm, and the numerical aperture 0.2. The output of the fiber was imaged in the crystal via two doublets, leading to a pump waist radius of about 230 micrometers in the crystal. With no amplification, the 8-pass system attenuated the input beam by only 15 %. We choose the focal length of the doublet so as to the numerical aperture of the cone formed by the eight beams passing through the crystal was slightly smaller than the numerical aperture of the pump beam, in order to optimize the overlap between the signal beam path and the volume where gain is available. The input beam size was adjusted in order to obtain a waist radius of 125 µm in the crystal.

## 3 Experimental results

### 3.1 Nd:YAG amplifier

The gain medium used for this experiment was an AR-coated composite Nd:YAG crystal formed by a 2 mm long undoped section and a 5 mm long 1.1% doped section. The square section is 2 x 2 mm$^2$. A dichroïc mirror (HT @ 808 nm, HR

@ 1064 nm) was coated on the rear face of the crystal (see figure 2). The crystal absorbed about 85 % of incident pump power.

We measured the extracted power - defined as the difference between the output and the input power - versus the pump power for a given injected power of 110 mW. The results were plotted on figure 3, where we can notice that saturation appears when pumping above 10W. This saturation is probably due to thermal lensing. In fact, the defocus induced by the thermal lens led to a non-optimal overlap between the pump and signal beam, and consequently reduced the efficiency of the system. Parasitic laser effect (between the prisms and the mirror coated on the crystal) has been observed and was considered after measurement as negligible (only 10 mW under amplifying conditions). Up-conversion processes or ASE losses are also probably involved in the degradation of the performance, but their influence is supposed to be very weak : the amplifier is saturated, so consequently all the physical effects linked to an important population in the upper level (parasitic laser effect, ASE and up-conversion) are reduced .

At a pump power of 13 W, the crystal was fractured. Therefore, the pump power was set to 12 W for the following experiments.

We plotted the extracted power versus the input power on the amplifier on figure 4. A maximum extracted power as high as 2.7 W was obtained for an input power of 110 mW. It is also interesting to observe the extracted power versus the gain as shown on figure 5. We can then deduce the small signal gain from the intercept between the curve and the horizontal axis [13] and we obtained for the 8 passes a small signal gain of 130 (or 21 dB). For the maximal extracted power (2.7 W) the gain is 21. Defining an extraction efficiency $\eta = (P_{out}-P_{in})/P_p$ where $P_{out}$, $P_{in}$ and $P_p$ are the output, input and pump power respectively, we find $\eta = 21$ % for an input power of 110 mW. The output beam was linearly polarized but the thermal lensing in the crystal induced a small degradation of the beam quality, leading to a $M^2$ of 1.8.

The observed saturation at high pump power, coupled with the degradation of the spatial quality of the output beam, limits the potential of this Nd:YAG based system for high power outputs. We consequently investigate another solution by using a low-doped Nd:YVO$_4$ crystal.

## 3.2 Nd:YVO4 amplifier

The very high emission cross-section of the Nd:YVO$_4$ makes this crystal very attractive for efficient amplification if we can solve the thermal problems induced by the low thermal conductivity of this crystal. We used a simple 10 mm long, 0.3 % doped crystal with AR coatings on the two faces. The conjugaison of low-doping and relatively long crystal allows a better distribution of the pump power inside the gain medium, leading to a weaker and less aberrating thermal lens. It also authorizes a higher pumping level (up to 15 W in our case) without the breaking problems often observed over 10W of pump power for 1% doped samples [14]. We also take benefit of the high-brightness of the pump diode to ensure a good overlap between pump and signal over the whole length of the crystal Low-doped Nd:YVO$_4$ crystals have often been studied in laser oscillators [15-16], but never as amplifiers to our knowledge. However, it is of particular interest in this last case to limit the up-conversion effects [17]. The population in the excited state is much greater in the amplifiers than in the oscillators, leading to a higher sensitivity to up-conversion trouble. A low doping level induces a decrease of the population density and consequently limits the up-conversion effects. A higher small signal gain is then expected with this low doping level. Because of the strong polarization-dependance of the Nd:YVO$_4$ gain, we could not use directly our prism-based system with this crystal. Indeed, each right-angle prism induced a rotation of the state of polarization of the beam, leading to a gain reduction for the most part of the passes. The solution to overcome this drawback is to control the state of polarization after each prism. To this end we put three small half-waves plates on the beam passes number 3,5, and 7 (figure 1b). Each of them can be individually rotated to restore the state of polarization corresponding to the highest emission cross section in Nd:YVO$_4$.

The 8-pass structure could not be studied because of strong parasitic laser effects between the prisms and the mirror. We then used our amplifier in a six-pass configuration when a parasitic laser power of only a few milliwatts was measured. The parasitic lasing effect was measured under amplifying and non-amplifying conditions. As the parasitic laser effect creates a beam larger than the amplified signal beam, it is possible to isolate the laser effect and to measure it in both situations. It appears that the relatively strong (80 mW) laser effect is reduced by a factor of 10 (becoming less than 8 mW) when amplifying : when an input signal is injected in the amplifier, the population of the upper level of the lasing transition decreases rapidly which leads to a weaker laser effect.

With this architecture the output power was 4.5 W with a total gain of 43 (16 dB) and an optical extraction efficiency of 33% (figures 3,4 and 5). Defining the small-signal gain as the extrapolated intercept with the horizontal axis on figure 4 [13], we obtained 42 dB (15800) of small signal gain. The amplifier shows therefore good efficiency for a large range of input power, from several microwatts to more than 100 mW. We do not observe any change of the state of polarization between input and output beams, and the amplified signal is nearly diffraction limited ($M^2_x$= 1.11 and $M^2_y$= 1.13), corresponding to the beam quality of the microlaser.

Although this half-waves plates system gave excellent results, it suffered from an increase of complexity and cost. Consequently we designed a much more simple, compact and low cost scheme by using corner cube retroreflectors instead of right angle prisms to obtain the lateral shift of the signal beam. Indeed, the silver coated corner cubes retroreflectors maintain polarization state through the different passes, which implies that the half waves plates are not useful anymore. However, the corner cube exhibits more losses than prisms (the transmission of a single corner cube is about 85%) and stronger parasitic laser effect because of lower alignement sensitivity. As a consequence, we found that the optimized configuration in terms of efficiency, compactness and simplicity used only one corner cube. In this case, the beam makes 4 passes in the gain medium in two planes of incidence (see figure 6).

With this simplified structure we measured 4 W of output power with a gain of 37, a small-signal gain as high as 15800 (42 dB) and an optical extraction efficiency of 27 % (figure 3,4 and 5). Although the alignment is slightly more complex, it is also possible to perform 6-pass configuration with two corner cubes. With this scheme (see fig 5), we obtained 5.7 W of output power for a 100 mW seed power and a small signal gain of more than 56 dB. The extraction efficiency reached 36 %. These results are better than those obtained with the previous configuration with the prisms and half-wave plates, because the use of corner cubes allows better overlap between the signal beams and the pump volume : indeed, the angle between the optical axis and the signal beams can be smaller with corner cubes than with the prism based system.

The typical shape of the input-output curve (figure 4) can be explained in terms of gain : for the Nd:YVO$_4$-based amplifier the small signal is as high as 9 per pass versus only 1.9 for Nd:YAG. When a small input power is seeded in the amplifier the non-saturated gain is very high (not far from $9^4$ for a 4-pass configuration) leading to a very strong slope in this regime. When the input signal becomes higher, the power achieved after several passes is high enough to strongly saturate the amplifier as shown theoretically by Frantz and Nodvik [18].

We also measure (Tektronix TDS 7254 oscilloscope) the temporal shape of the pulse before entering the amplifier system and after amplification (figure 7). The result is a shortening of the pulse duration (from 700 ps to about 500 ps) due to the lower gain observed for the end of the pulse caused by gain saturation.

The $M^2$ factor was found to be 1.15 in the two directions and the output was lineary polarized for all the configurations.

We then performed frequency doubling and tripling of the output of our amplifier. With 4 W of 1064 nm radiation incident on lithium triborate crystals (due to unoptimized design), 1.72 W of second harmonic at 532 nm (43 % of conversion efficiency) and 650 mW of third harmonic at 355 nm were generated.

## 4 Conclusion

In conclusion, we demonstrate very simple, compact and low cost longitudinally pumped multipass amplifiers. Thanks to the use of a new tridimensionnal structure based on right angle prisms and a low-doped Nd:YVO$_4$ crystal, we obtained a 4.5W diffraction-limited output with a gain of 43 and an extraction efficiency of 33%. Another structure based on corner cubes retroreflectors and using the same crystal is also described : the simplicity and compactness of the system are increased and the performance is strongly enhanced : we reached 5.7W of extracted power in a 6-pass structure with a gain of 57 (17.5 dB) and a small signal gain of 56 dB. Moreover, the amplifier exhibits efficient behaviour for a large range of input powers. The performance of this amplifier present to our knowledge the best compromise between gain and extraction efficiency among the previously published comparable systems. This amplifier could also be seen as an interesting mix between the advantages of bulk crystals systems supporting high peak power pulses and of fiber based amplifier systems for its compactness and high small signal gain.

FIGURE CAPTIONS :

Figure 1 : (a) 3D representation of the circular geometry of the eight passes. M stands for the dichroïc mirror coated on the rear face of the crystal, L is the focusing doublet. (b) Front view seen from the crystal. Arrows describe the lateral shift induced by the prisms. The beam passes (1-2), (3-4), (5-6) and (7-8) define the four planes of incidence (broken lines).

Figure 2 : schematic of the experimental setup and the Nd:YAG composite crystal (in the middle).

Figure 3 : extracted power (defined as the difference between output and input power) versus the pump power for the Nd:YAG crystal and the Nd:YVO$_4$ crystal in two different configurations (see text).

Figure 4 : extracted power as a fonction of the input power for the Nd:YAG crystal and the Nd:YVO$_4$ crystal in three different configurations (see text).

Figure 5 : extracted power versus the gain in decibels for the Nd:YAG crystal and the Nd:YVO$_4$ crystal in three different configurations (see text).

Figure 6 : amplifier scheme with the corner cube retroreflector configuration. Lenses L1 and L2 are used to image the pump in the crystal. The mirror M is AR coated @ 808 nm and HR @ 1064 nm. A single corner cube is used to perform 4 passes amplification. Bottom right, a 2D front view that shows the two planes of incidence (broken lines) used in this structure.

Figure 7 : temporal shape of the pulses before and after amplification.

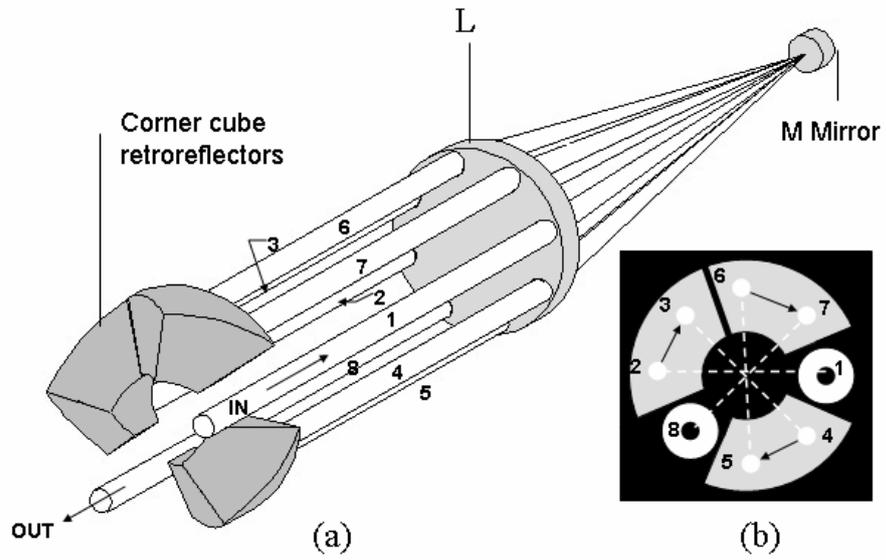

Figure 1 : (a) 3D representation of the circular geometry of the eight passes. M stands for the dichroïc mirror coated on the rear face of the crystal, L is the focusing doublet. (b) Front view seen from the crystal. Arrows describe the lateral shift induced by the prisms. The beam passes (1-2), (3-4), (5-6) and (7-8) define the four planes of incidence (broken lines).

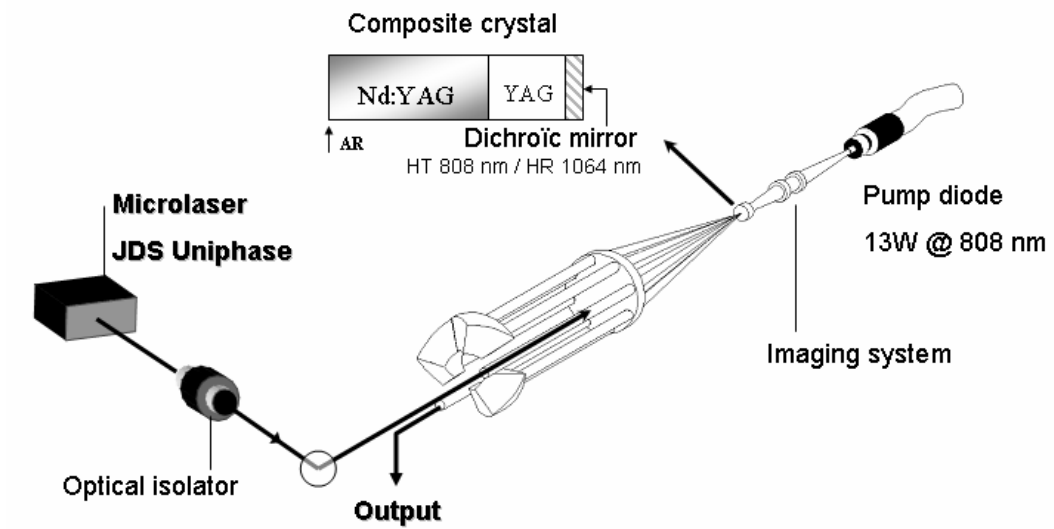

Figure 2 : schematic of the experimental setup and the Nd:YAG composite crystal (in the middle).

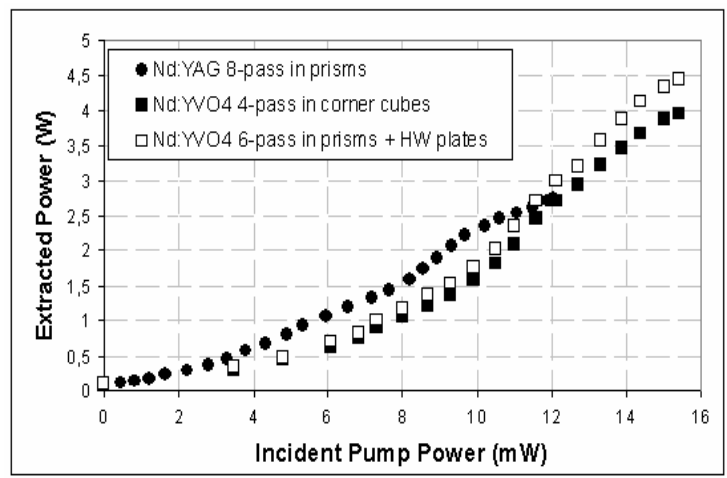

Figure 3 : extracted power (defined as the difference between output and input power) versus the pump power for the Nd:YAG crystal and the Nd:YVO$_4$ crystal in two different configurations (see text).

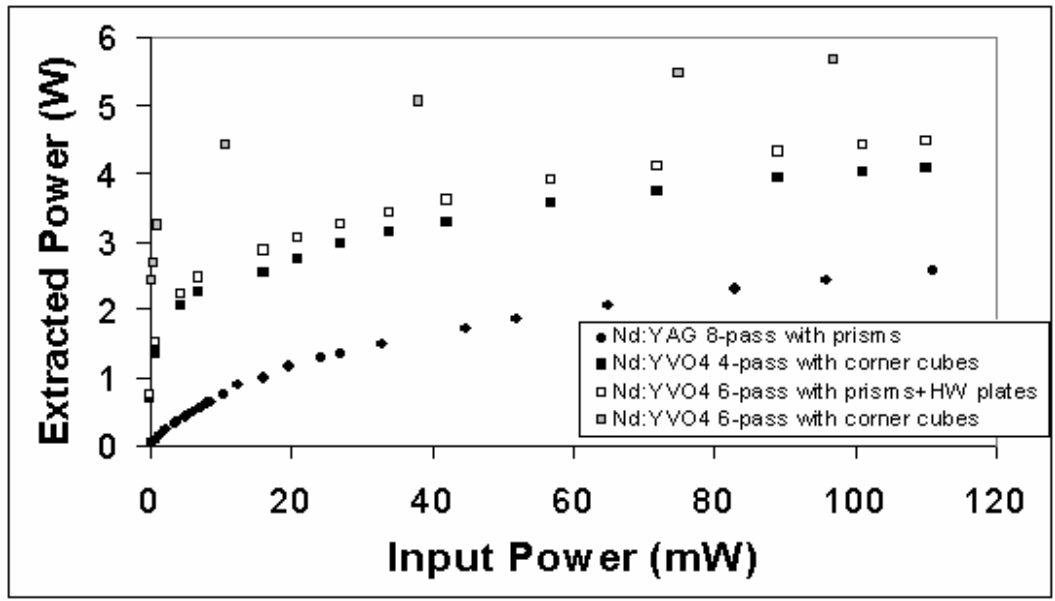

Figure 4 : extracted power as a fonction of the input power for the Nd:YAG crystal and the Nd:YVO$_4$ crystal in three different configurations (see text).

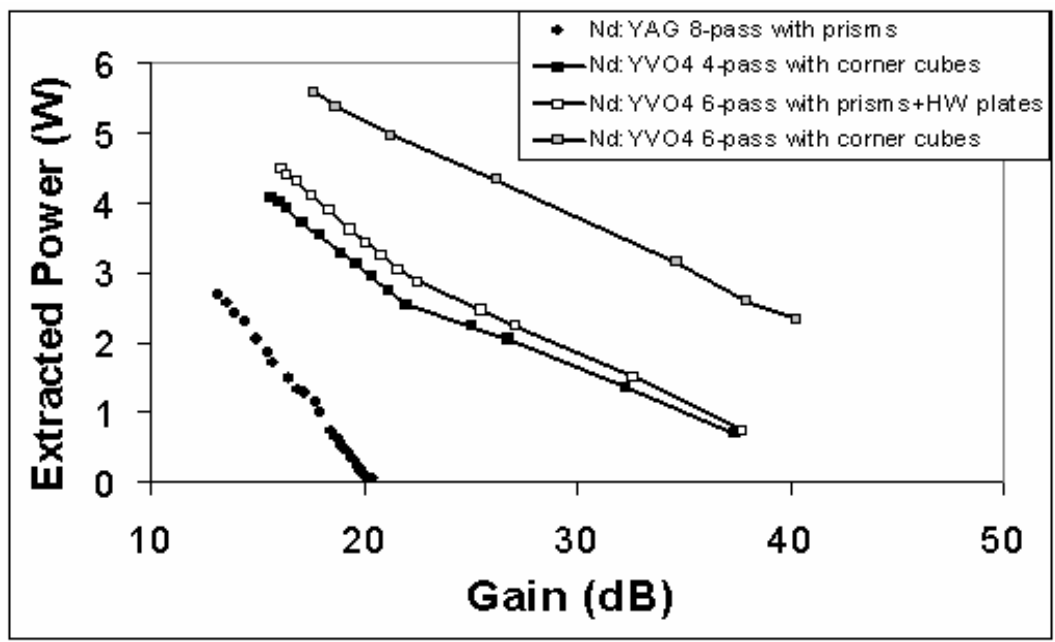

Figure 5 : extracted power versus the gain in decibels for the Nd:YAG crystal and the Nd:YVO$_4$ crystal in three different configurations (see text).

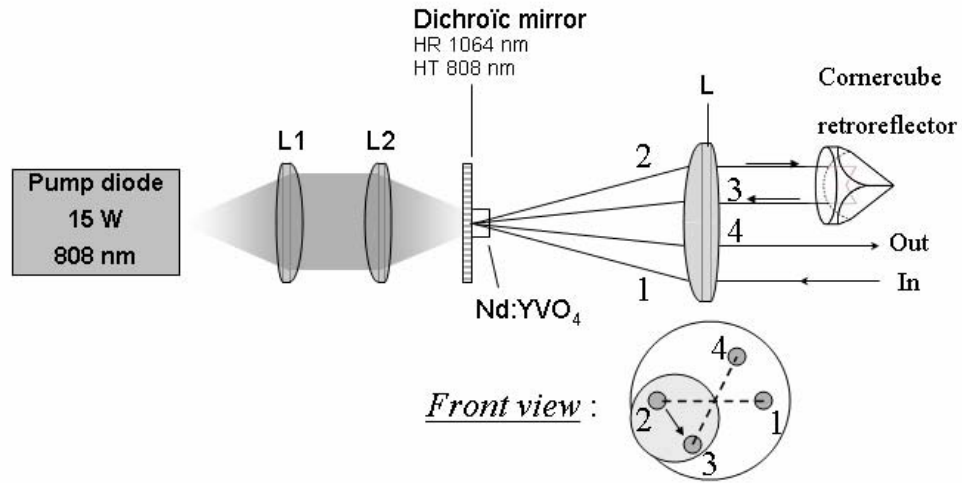

Figure 6 : amplifier scheme with the corner cube retroreflector configuration. Lenses L1 and L2 are used to image the pump in the crystal. The mirror M is AR coated @ 808 nm and HR @ 1064 nm. A single corner cube is used to perform 4 passes amplification. Bottom right, a 2D front view that shows the two planes of incidence (broken lines) used in this structure.

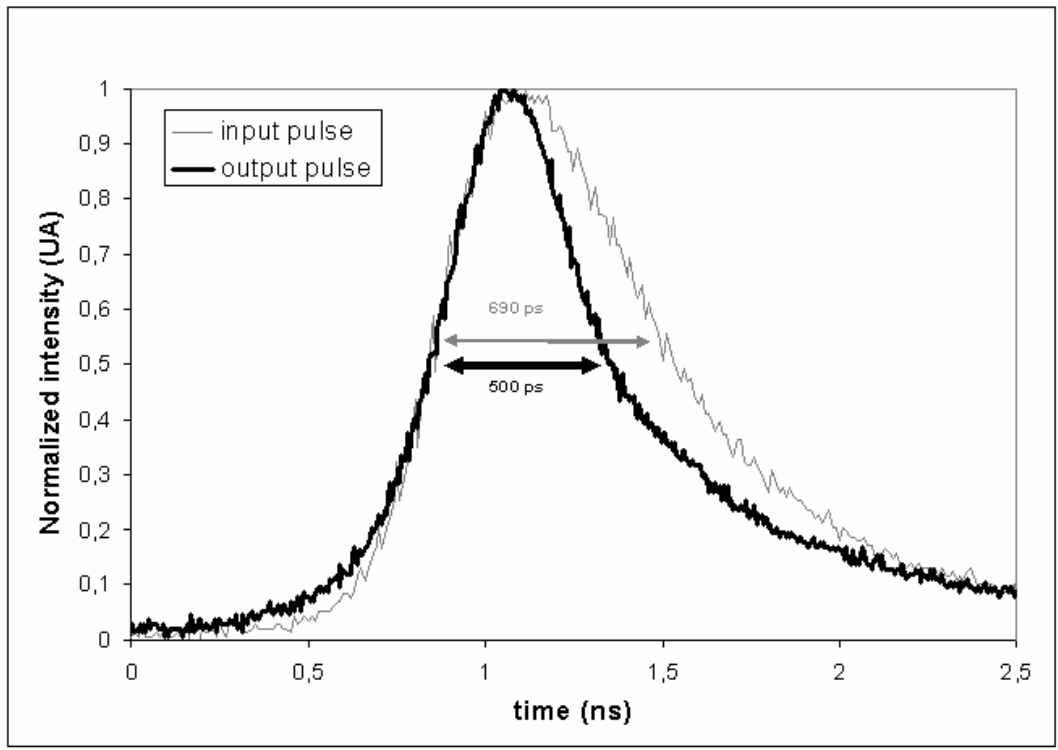

Figure 7 : temporal shape of the pulses before and after amplification.